\documentclass[conference]{IEEEtran}
\IEEEoverridecommandlockouts
\usepackage{cite}
\usepackage{amsmath,amssymb,amsfonts}
\usepackage{algorithmic}
\usepackage{graphicx}
\usepackage{textcomp}
\usepackage{color}
\usepackage{tabu}
\usepackage{multirow}
\usepackage{cite}
\usepackage{amsmath,amssymb,amsfonts}
\usepackage{algorithmic}
\usepackage{graphicx}
\usepackage{textcomp}
\usepackage{xcolor}
\usepackage[linesnumbered,ruled,vlined]{algorithm2e}
\usepackage{amsmath,amsfonts}
\usepackage{array}
\usepackage[caption=false,font=normalsize,labelfont=sf,textfont=sf]{subfig}
\usepackage{textcomp}
\usepackage{url}
\usepackage{verbatim}
\usepackage{graphicx}
\usepackage{cite}
\usepackage{pifont}
\usepackage[linesnumbered,ruled,vlined]{algorithm2e}
\usepackage{color}
\usepackage{booktabs}
\usepackage{tabu}
\usepackage{multirow}
\usepackage{graphicx}
\usepackage{float}
\usepackage{subfig}
\usepackage{makecell}
\usepackage{marvosym}
\usepackage{geometry}
\def\BibTeX{{\rm B\kern-.05em{\sc i\kern-.025em b}\kern-.08em
    T\kern-.1667em\lower.7ex\hbox{E}\kern-.125emX}}
\begin{document}

\title{Facing Unknown: Open-World Encrypted Traffic Classification Based on Contrastive Pre-Training}
\author{
  \IEEEauthorblockN{
    Xiang~Li\IEEEauthorrefmark{1}\IEEEauthorrefmark{2},
    Beibei~Feng\IEEEauthorrefmark{1}\IEEEauthorrefmark{2},
    Tianning~Zang\IEEEauthorrefmark{1}\IEEEauthorrefmark{2}$^{(\textrm{\Letter})}$,
    Shuyuan~Zhao\IEEEauthorrefmark{1},
    and Jingrun~Ma\IEEEauthorrefmark{1}\IEEEauthorrefmark{2}
  }
  \IEEEauthorblockA{\IEEEauthorrefmark{1} Institute of Information Engineering, Chinese Academy of Sciences, Beijing, China}
  \IEEEauthorblockA{\IEEEauthorrefmark{2} School of Cyber Security, University of Chinese Academy of Sciences, Beijing, China}
  \IEEEauthorblockA{\IEEEauthorrefmark{3} National Computer Network Emergency Response Technical Team/Coordination Center of China, Beijing, China}
  {Email: \{lixiang1, fengbeibei, zangtianning, zhaoshuyuan, majingrun\}@iie.ac.cn}
}

\maketitle

\begin{abstract}
Traditional Encrypted Traffic Classification (ETC) methods face a significant challenge in classifying large volumes of encrypted traffic in the open-world assumption, \emph{i.e.}, simultaneously classifying the known applications and detecting unknown applications. We propose a novel Open-World Contrastive Pre-training (OWCP) framework for this. OWCP performs contrastive pre-training to obtain a robust feature representation. Based on this, we determine the spherical mapping space to find the marginal flows for each known class, which are used to train GANs to synthesize new flows similar to the known parts but do not belong to any class. These synthetic flows are assigned to Softmax's unknown node to modify the classifier, effectively enhancing sensitivity towards known flows and significantly suppressing unknown ones. Extensive experiments on three datasets show that OWCP significantly outperforms existing ETC and generic open-world classification methods. Furthermore, we conduct comprehensive ablation studies and sensitivity analyses to validate each integral component of OWCP.

\end{abstract}


\section{Introduction}
Traffic classification, which groups similar or related traffic data by protocol or source, is crucial for ensuring 
Quality of Service (QoS), Quality of Experience (QoE), 
network management, Web measurement, and threat detection \cite{sirinam2018deep}. In recent years, we have witnessed the rapid development of new network technologies and mobile ecosystems, accompanied by one key evolution, \emph{i.e.}, transforming traffic from plaintext to encrypted form. Many websites and mobile applications (apps) now use Transport Layer Security (TLS) to protect privacy and ensure secure communication. According to the Annual Report of Let's Encrypt \cite{letsencryptorg}, HTTPS page loads have reached 84\% globally. Unfortunately, even malware apps have started using TLS to conceal communications for Command and Control (C\&C) and data theft. As reported by Sophos \cite{sophos}, encrypted traffic now accounts for 46\% of all malware. 
Large amounts of encrypted traffic, particularly those of unknown classes, present a new challenge to traditional Encrypted Traffic Classification (ETC) methods.

\begin{figure}[]
\centering
\includegraphics[scale=0.65]{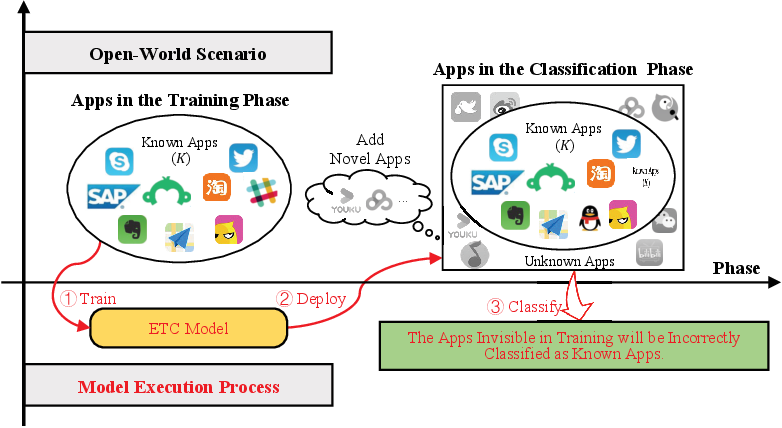}
\caption{An ETC Model Trained in the Closed-World Scenario cannot Handle the Open-World Classification Task.}
\label{fig1}
\end{figure}

Most of the existing ETC methods \cite{liu2019FS, he2020pert} operate under the closed-world assumption, which means that the apps presented in the classification phase must also be present during the model training phase.
If an app is invisible during training, it will be misclassified as a known app, as shown in Fig. \ref{fig1}. Enumerating all the apps and collecting their traffic for model training is impossible, as Google Play Store has over 2.6 million available apps \cite{statista}. To make matters worse, recent research indicates that app developers widely use Third-Party Libraries (TPLs), leading to homogeneous network behavior \cite{van2020flowprint}. Many app developers use standard repositories and associated domains to implement functions such as authentication, advertising, and analytics, which may increase the risk of false positives. 
For example, when \emph{TaoBao} app runs, not all traffic is directly related to itself. Some actions lead to \emph{*.amap.com} or \emph{*.alicdn.com}, which may also appear when running new apps, such as \emph{Youku} app. Moreover, over 3.17 million new malware apps are discovered yearly, many of which do not belong to any known app family \cite{gulyas2021cyber}. Traditional ETC methods fail to handle the open-world classification task that classifies known and detects unknown apps simultaneously.

One potential solution to address the challenge of detecting unknown apps is based on Softmax output, which provides a predicted probability distribution of a multiclass classification model. If all the dimensional values of the distribution are lower than a threshold probability, then it is considered an unknown class. This approach has proven successful in Computer Vision field \cite{bendale2016towards, yoshihashi2019classification}. However, encrypted traffic presents unique challenges compared to images, such as unreadability and homogeneity to result in significant overlap between apps, which makes the framework inappropriate for ETC. 

To tackle the issue, we propose OWCP, a novel Open-World Contrastive Pre-training framework for ETC that addresses the limitations of existing methods. OWCP provides a robust feature representation by adopting traffic contrastive  pre-training. Specifically, we resample the training data by constructing positive and negative flow pairs, bringing positive pairs closer together and pushing negative ones farther to optimize the discriminable feature representation. On this basis, we determine the spherical mapping space for each known class to find non-homogeneous flows at the margins, which are used to train Generative Adversarial Networks (GANs) to synthesize new flows that are similar to known flows but do not belong to any known class. The above operations bring two salient advantages: i) any flow that deviates from the distribution of known app classes can be classified as unknown app flow, so that synthesizing marginal flows can approach the known while falling into the simulated unknown; ii) by filtering out homogeneous marginal flows that are generated by reusing TPLs, we can focus on unknown flows triggered by new apps of their own behavior. Then, the synthetic flows are assigned to unknown nodes to modify the classifier, to overall reduce and flatten the recognition probability of unknown flows, making the classifier more sensitive to known flows and significantly suppressing unknown ones.

Our major contributions can be concluded as follows:
\begin{itemize}
    \item We propose a novel open-world ETC paradigm that addresses the challenge of unknown classes in real network environments. To the best of our knowledge, OWCP is the first method that uses pre-training to solve the open-world ETC task.  
    
    \item OWCP discovers non-homogeneous marginal flows in spherical space and synthesizes simulated unknown flows to jointly improve the performance of known classification and unknown detection.
    
    \item We evaluate the proposed framework in extensive experiments across three public available datasets and demonstrate its superiority over existing ETC models and multiple open-world classification methods. 
\end{itemize}

\section{Related work}
As machine learning and deep learning become more widely adopted, researchers have increasingly focused on developing ETC methods and have reported high accuracy. In this section, we will discuss various classification models that rely on different types of features, which can be divided into {statistical-feature-based methods} and {sequence-feature-based methods}.

\subsection{Statistical-Feature-Based Methods}
Early on, a combination of various traditional machine learning algorithms proposed statistical features to solve ETC tasks. Representatively, Taylor \emph{et al.} \cite{taylor2017robust} first designed the burst and flow statistical features and offered a robust app classification method. Recent statistical features captured complex dependencies through deep learning. Shi \emph{et al.} \cite{shi2018efficient} built a deep learning framework to select and combine the statistical features to enhance the performance of traffic classification. Chen \emph{et al.} \cite{chen2019rethinking} exploited attribute features and statistical features to predict the app to which the encrypted flow belongs. 
However, these methods are mainly based on rich experiences, professional knowledge, and much human effort. 

\subsection{Sequence-Feature-Based Methods}
Encryption apps leak information about dependencies or transfers between data messages, known as sequence features.
Korczynski \emph{et al.} \cite{2014Markov} proposed representing the message type sequence of TLS handshake phase data with a Markov transformation matrix to classify encrypted traffic. 
Fu \emph{et al.} \cite{fu2016service} extracted packet length and time delay sequences to build a Hidden Markov Model, which in turn enhances the intrinsic richness of the model. 
Meanwhile, researchers have started to explore convolutional neural networks and recurrent neural networks combined with the non-plaintext payload sequence. 
Wang \emph{et al.} \cite{wang2017hast} used the first 784 bytes of the payloads to construct an end-to-end classification model. 
Lotfollahi \emph{et al.} \cite{lotfollahi2020deep} proposed the Deep packet method and adopted the stacked autoencoder to extract features from encrypted payloads.

Most existing ETC classification methods have primarily relied on a closed-world assumption, meaning they can only classify traffic within a static dataset of pre-defined app classes. In open-world assumptions, these classifiers must be able to detect unknown classes to provide accurate traffic classification to support network measurement and management.


\section{The Proposed OWCP}
\begin{figure}[!ht]
\centering
\includegraphics[scale=0.65]{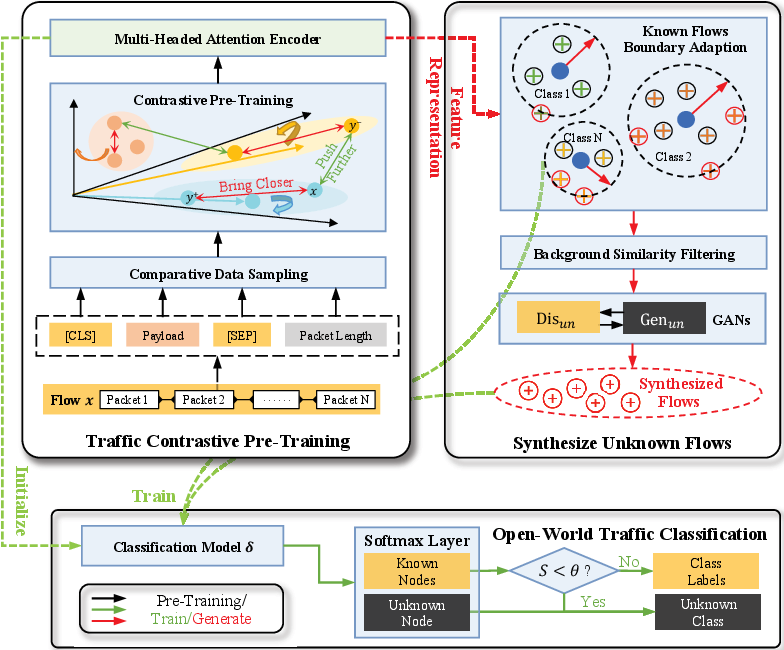}
\caption{Overview of OWCP workflow.}
\label{fig2}
\end{figure}

This section provides a formal problem definition for the open-world ETC task and details our proposed OWCP framework. As shown in Fig. \ref{fig2}, we first obtain a pre-trained model using contrastive learning to identify non-homogeneous flows at the margins by mapping them onto a spherical space. A GANs generator is then used to synthesize the distribution of marginal non-homogeneous flows to simulated unknown flows that support the classifier's training. By adding an unknown node in Softmax, we can improve the performance of classification and detection by lowering and flattening the recognition probabilities of unknown flows while relatively raising those of known flows.

\subsection{Problem Definition}
Give a set of labeled known app flows $\mathcal{D}=\{(x_0, l_0),$ $(x_1, l_1), \ldots, (x_n, l_n)\}$, 
where $x_n$ is a flow instance containing non-plaintext payload and packet length sequence features, with app label $l_n$ belong to $K=\{0,1,...,k\}$ known apps. A DL-based classifier $F\left(x_i\right) \rightarrow \hat{l}_i$ is created from $\mathcal{D}$ to predict label $\hat{l}_i$ of $x_i$ that matches the actual label $l_i$. $F\left(x_i\right)$ aims to classify test flows from not only known apps $K$ but also unknown classes $K^u$, where $K^u \cap K=\phi$. 

\begin{figure}[!ht]
\centering
\includegraphics[scale=0.53]{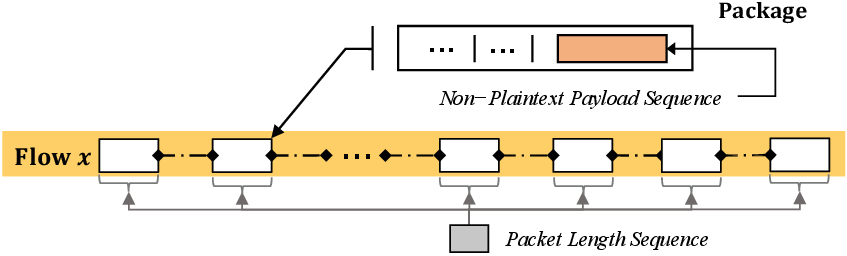}
\caption{Non-Plaintext Payload and Packet Length Sequence Features.}
\label{fig3}
\end{figure}

Meanwhile, we define the non-plaintext payload and packet length sequence features, which are illustrated in Fig. \ref{fig3}. 

\subsubsection{Non-Plaintext Payload Sequence}
Although TLS encrypts the plaintext, some side-channel information is still leaked from the encrypted payload. To obtain more context-sensitive information, we partition the payload by two bytes, \emph{e.g.}, from \{1a, 2b, 03, 45, 62, aa, ...\} to \{1a2b, 0345,62aa, ...\}.

\subsubsection{Packet Length Sequence}
The packet length is measured in bytes, and we use “+” to indicate packets sent from the client to the server and “-” to indicate packets sent from the server to the client, \emph{e.g.},\{+328, -1074, -180, +328...\}, to represent the bi-directional nature of the flow. 

\subsection{Traffic Contrastive pre-training}
Without loss of generality, we assume $x$ is one of the bidirectional flows of the training set, which is obtained according to the traffic five-tuple. 
We filter out biased information for each packet, such as IP address, port number, MAC address, \emph{etc}. 
Based on existing works \cite{lotfollahi2020deep}, we extract the $M*64$ bytes non-plaintext payload sequence of $M$=6 packets and the packet length sequence of $N$=128 packets, reconstructed and encoded into the coding dictionary by order of frequency of occurrence. Four unique markers, CLS, SEP, PAD, and UNK, are added to the dictionary, indicating the start flag of the input, the separator flag of the non-plaintext payload sequence and the packet length sequence, the padding flag, and the unregistered word flag, respectively. Let the Non-plaintext Payload (NP) and Packet Length (PL) sequences after dictionary encoding as $NP=\left[np_0, \ldots, np_M\right]$ and $PL=\left[pl_0, \ldots, pl_N\right]$, and the input sequence of x can be expressed as, 
\begin{equation}
x=CLS+[NP]+SEP+[PL]
\end{equation}

To capture more internal connections, we apply word embedding to $x$ using the parameter matrix $\mathrm{W} \in \mathbb{R}^{V \times d}$, which transforms the discrete input sequence $x \in \mathbb{R}^{(m * 64 / 2+n+2) \times 1}$ into a high-dimensional vector $x \in \mathbb{R}^{(m * 64 / 2+n+2) \times d}$, where $V$ denotes the size of $\mathrm{W}$, and $d$ is the dimension. Additionally, we incorporate positional encoding information into the embedded vector to enhance its contextual representation \cite{vaswani2017attention}. 

\begin{figure}[!ht]
\centering
\includegraphics[scale=0.75]{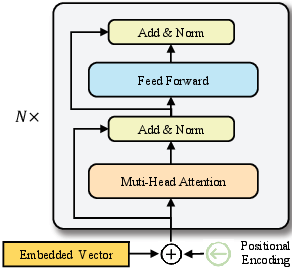}
\caption{Multi-Headed Attention Encoder Structure. Add \& Norm Means the Residual Network and Layer Normalization \cite{vaswani2017attention}.}
\label{fig4}
\end{figure}

OWCP resamples the training data by constructing positive and negative flow pairs to obtain the pre-training set. Specifically, we randomly select samples $y^+$ and $y^-$, one from the same class as the positive example and the other from a different class as the negative example, respectively. Then the triple [$x, y^+, y^-$] feeds to the multi-headed attention encoder, consisting of multi-headed self-attention and feedforward neural networks, as shown in Fig. \ref{fig4}, which can be stacked $N$=6 times to enhance the representation capability. The multi-headed self-attention network encodes contextual information, and the feedforward neural network provides nonlinear variation, consisting of two linear layers and one layer of ReLU, which is connected by residual networks and layer normalization. Meanwhile, we use the InfoNce loss function to pre-train the multi-headed attention encoder in order to bring the positive example closer and push the negative example further, calculated as follows,
\begin{equation}
\begin{aligned}
L_{C T L}=-\log \frac{\exp \left(\operatorname{sim}\left(x, y_i^{+}\right) / \tau\right)}{\sum_{k=1}^N \exp \left(\operatorname{sim}\left(x, y_k^{+}\right)\right)}
\\
-\log \frac{\exp \left(\operatorname{sim}\left(x, y_i^{-}\right) / \tau\right)}{\sum_{k=1}^N \exp \left(\operatorname{sim}\left(x, y_k^{-}\right)\right)}
\end{aligned}
\end{equation}

where $\tau$ denotes the temperature parameter used to control the shape of the distribution of logits. 

\subsection{Synthesize Unknown Flows}
Theoretically, any flow that deviates from the distribution of known app classes can be classified as unknown app flow \cite{guo2021conservative}. 
We first discover the marginal flows with discrimination attributes in each known class by spherical decision boundaries. Specifically, we use the pre-trained multi-headed attention encoder $\theta$ as the feature extractor for known app classes, to determine the spherical center by computing the average feature vector for each class. Let $A_k$ denote the set of flows labeled with class $k$. The centroid $c_k$ is the mean vector of flows in $A_k$, 
\begin{equation}
\boldsymbol{c}_k=\frac{1}{\left|A_k\right|} \sum_{\left(x_i, l_i\right) \in A_k}{\theta\left(\boldsymbol{x}_i\right)}
\end{equation}

where $\left|A_k\right|$ denotes the number of flows in $A_k$. We define $\Delta_k$ as the radius of decision boundary with respect to the centroid. For each flow of $A_k$, we aim to satisfy the constraint,
\begin{equation}
\forall \boldsymbol{x}_i \in A_k,\left\|\theta\left(\boldsymbol{x}_i\right)-\boldsymbol{c}_k\right\|_2 \leq \Delta_k
\end{equation}

where $\left\|\theta\left(\boldsymbol{x}_i\right)-\boldsymbol{c}_k\right\|_2$ denotes the Euclidean distance between $\theta\left(\boldsymbol{x}_i\right)$ and $c_k$. 
We collect the marginal flow set $\mathcal{D}^{'}=\{x_0, \ldots, x_m\}$, where each instance is located/close to the margin of $\Delta$, \emph{i.e.}, $(\Delta-\left\|\theta\left(\boldsymbol{x}_i\right)-\boldsymbol{c}_k\right\|_2) < \varepsilon$, and $m$ \textless\textless $n$. 
We consider the existence of homogeneous marginal flows in $\mathcal{D}^{'}$ generated by the reuse of TPLs, inconsistent with the goal of focusing on the unknown apps themselves. For that, we propose to filter homogeneous marginal flow in $\mathcal{D}^{'}$ based on background similarity if the following conditions are satisfied:
\begin{itemize}
    \item The \{destination IP address, destination port\} tuple of the marginal flow appears in multiple classes.  
    \item The \{SNI\} or \{TLS Certificate\} of the marginal flow appears in multiple classes. 
\end{itemize}


Inspired by GANs \cite{goodfellow2020generative}, flow generation is achieved without explicitly modeling probability densities.
We use a generator $Gen$ for sampling a latent variable $z$ from a prior distribution, \emph{e.g.}, a Gaussian $\mathcal{N}$ as the input to generate an output $Gen(z)$. Meanwhile, a discriminator $Dis$ is trained to distinguish whether an input $x$ is from a target data distribution by mapping ${x} \in \mathcal{D}^{\prime} \ or \ {Gen}({z})$ to a probability range in [0,1]. $Gen$ aims at synthesizing simulated flows as accurately as possible when freezing $Dis$, while $Dis$ aims to distinguish when freezing $Gen$,  which are contesting with each other in a zero-sum game framework. The generation of simulated unknown flows $\mathcal{D}^u=\{Gen(z)\}$ can be optimized by a min-max objective of compact form as, 
\begin{equation}
\min_{Gen} \max_{Dis} \mathbb{E}_{x \in \mathcal{D}^{\prime}}[\log Dis(x)]+\mathbb{E}_{z \in \mathcal{N}}[\log (1-Dis(Gen(z)))]
\end{equation}

\subsection{Open-world Traffic Classification}
We construct the classification model $\delta$, which adds a fully connected layer adapted to the classification task, initialized by the parameters of the pre-trained $\theta$ and fine-tuned by the $\mathcal{D}$ and $\mathcal{D}^u$. 
Correspondingly, we add an unknown decision node in the Softmax layer to implicitly modify the $\delta$ by overall reducing and flattening the recognition probability of unknown flows, to relatively sensitize to known flows while significantly suppressing unknown ones. We use a modified two-stage recognition to provide the final classification results as follows,
\begin{equation}
\hat{l} = \begin{cases}\operatorname{argmax}_{l \in\{0, \ldots, k\}} P(l \mid \delta(x)) & \text { if } P(l \mid \delta(x)) \geq \sigma \\ {unknown} & \text { otherwise. }\end{cases}
\end{equation}

where $P(l \mid \delta(x))$ is the output of the SoftMax layer, and $\sigma$ is a hyperparameter threshold that can be selected by doing a grid search calibration procedure using a set of training flows plus a sampling of open-world flows. 

\section{Experiment}
\subsection{Experiment Setting}
\subsubsection{Dataset Description}
To comprehensively evaluate the performance of our proposed OWCP, we conduct experiments on three public available datasets:
\begin{itemize}
    \item CrossPlatform \cite{ren2019international} consists of traffic from popular apps on the Android platform in China, United States, and India, with a total of 215 classes.  
    \item ISCX17 \cite{draper2016characterization} includes 7 types of traffic for VPN and non-VPN communications, combined by apps, resulting in 17 different classes. 
    \item USTC-TFC \cite{wang2017malware} contains 10 classes of benign traffic and 10 classes of malicious traffic. 
\end{itemize}

We randomly select 80\% of the apps from each dataset as known classes, while the remaining apps are treated as unknown classes. The data from known classes are randomly divided into training, validation, and test sets, with a ratio of 8:1:1. 
Among them, the test set of the known classes is used as CW-test set.
All data from unknown classes are added to the CW-test set to form the OW-test set.
The detailed dataset settings are provided in Table \ref{Tab1}.

\begin{table}[htbp]
\captionsetup{font={footnotesize}}
\renewcommand\arraystretch{0.8}
\scriptsize
\caption{The Statistical Information of the Datasets. CW and OW mean the close-world and open-world, respectively.}
\centering
\resizebox{\linewidth}{!}{
\begin{tabular}{c|cc|ccc}
\toprule
\thead{\textbf{Dataset}}  
& \thead{\textbf{Known}    \\  \textbf{Class}}
& \thead{\textbf{Unknown}  \\  \textbf{Class}}
& \thead{\textbf{Training}  \\  \textbf{Flows}}
& \thead{\textbf{CW-Test} \\  \textbf{Flows}}
& \thead{\textbf{OW-Test} \\  \textbf{Flows}}
\\ \cmidrule{1-6}
{CrossPlatform} & 172 & 43 & 19,117 & 2,389 & 6,343 
\\ \cmidrule{1-6}
{ISCX17} & 14 & 3 & 1544 & 197  & 568
\\ \cmidrule{1-6}
{USTC-TFC} & 16 & 4 & 40,560 & 50,68 & 13,037 
\\ \bottomrule
\end{tabular}}
\label{Tab1}
\end{table}

\subsubsection{Implementation Details and Evaluation Metrics}
All experiments are conducted on a server with 128GB RAM, Intel(R) i7-8700 CPU, and NVIDIA 3090 GPUs, implemented with Pytorch 1.10.0. 
The multi-head attention encoder contains 8 heads, with a vector dimension of 64 for $q_i$, $k_i$, $v_i$, and the feedforward neural network has 1024 neurons. We use the BertAdam optimizer with a learning rate of 5e-5 and a warmup of 0.03.
To select the hyperparameter $\sigma$, we perform a grid search calibration procedure and cross-validation (0.7 in this paper).
We evaluate and compare the performance by the closed-world metrics, including Accuracy (AC) and
F1 score \cite{liu2019FS}. 
Following \cite{bendale2016towards, yoshihashi2019classification}, we also use the open-world metrics to measure the evaluation performance (AC${}_{ow}$, F1${}_{ow}$), \emph{e.g.}, AC${}_{ow}$ is defined as,
\begin{equation}
{AC}_{ow}=(\frac{\mathrm{TP}_{(K)}}{\mathrm{TP}_{(K)}+\mathrm{FN}_{(K)}}+\frac{\mathrm{TP}_{(U)}}{\mathrm{TP}_{(U)}+\mathrm{FN}_{(U)}})/2
\end{equation}

where ${TP}_{(K/U)}$ and ${FN}_{(K/U)}$ are the true positive and the false negative for known/unknown classes, respectively. Macro Average \cite{van2020flowprint} is used to avoid biased results due to imbalance between multiple classes of data by calculating the mean.  

\subsection{Comparison with Existing Methods}
To get a comprehensive understanding of the OWCP performance, we compare with four ETC methods and two generic open-world classification methods: 
\begin{itemize}
    \item Deep Fingerprinting (DF) \cite{sirinam2018deep} and FlowPrint \cite{van2020flowprint}, which both support open-world ETC. 
    \item Fs-Net \cite{liu2019FS} and PERT \cite{he2020pert} (with pre-training), which are closed-world ETC methods.
    \item Pre-trained encoders equipped with generic open-world classification methods, namely PT-OM (with Openmax \cite{bendale2016towards}) and PT-ST (with thresholding Softmax \cite{yoshihashi2019classification}), which are fine-tuned only by the training set.
\end{itemize}

\begin{table}[!]
\centering
\scriptsize
\captionsetup{font={footnotesize}}
\caption{Close-world Comparison Results on CrossPlatform, ISCX17, USTC-TFC.}
\label{Tab2}
\renewcommand\arraystretch{0.8}
\resizebox{\linewidth}{!}{
\begin{tabu}{l|cc|cc|cc}
\toprule
\multirow{2}{*}{\textbf{Method}}
& \multicolumn{2}{c|}{\textbf{CrossPlatform}} 
& \multicolumn{2}{c|}{\textbf{ISCX17}}
& \multicolumn{2}{c}{\textbf{USTC-TFC}}
\\ \cmidrule(r){2-3} \cmidrule(r){4-5} \cmidrule(r){6-7}
& AC    & F1
& AC    & F1
& AC    & F1
\\ \cmidrule(r){1-7} 
{DF\cite{sirinam2018deep}}
& 47.86 & 39.67 
& 65.89 & 57.99
& 82.49 & 80.67 
\\ \cmidrule(r){1-1} \cmidrule(r){2-3} \cmidrule(r){4-5} \cmidrule(r){6-7}
{FlowPrint\cite{van2020flowprint}}
& \textcolor{white}{x}90.77\textcolor{white}{x} & \textcolor{white}{x}92.36\textcolor{white}{x} 
& \textcolor{white}{x}\textbf{88.73}\textcolor{white}{x} & \textcolor{white}{x}69.24\textcolor{white}{x} 
& \textcolor{white}{x}85.81\textcolor{white}{x} & \textcolor{white}{x}75.26\textcolor{white}{x}
\\ \cmidrule(r){1-1} \cmidrule(r){2-3} \cmidrule(r){4-5} \cmidrule(r){6-7}
{Fs-Net\cite{liu2019FS}}
& 54.73 & 48.26 
& 70.67 & 52.48 
& 90.47 & 89.86
\\ \cmidrule(r){1-1} \cmidrule(r){2-3} \cmidrule(r){4-5} \cmidrule(r){6-7}
{PERT\cite{he2020pert}}
& 97.89 & 89.14 
& 83.18 & 71.23 
& 99.37 & 99.40 
\\ \cmidrule(r){1-1} \cmidrule(r){2-3} \cmidrule(r){4-5} \cmidrule(r){6-7}
\textbf{OWCP}
& \textbf{98.88} & \textbf{96.21} 
& 86.79 & \textbf{79.02}
& \textbf{99.85} & \textbf{99.56}
\\ \bottomrule
\end{tabu}}
\end{table}

\begin{table}[!]
\centering
\scriptsize
\captionsetup{font={footnotesize}}
\caption{Open-world Comparison Results on CrossPlatform, ISCX17, USTC-TFC.}
\label{Tab3}
\renewcommand\arraystretch{0.8}
\resizebox{\linewidth}{!}{
\begin{tabu}{l|cc|cc|cc}
\toprule
\multirow{2}{*}{\textbf{Method}}
& \multicolumn{2}{c|}{\textbf{CrossPlatform}} 
& \multicolumn{2}{c|}{\textbf{ISCX17}}
& \multicolumn{2}{c}{\textbf{USTC-TFC}}
\\ \cmidrule(r){2-3} \cmidrule(r){4-5} \cmidrule(r){6-7}
& AC${_{ow}}$    & F1${_{ow}}$
& AC${_{ow}}$    & F1${_{ow}}$
& AC${_{ow}}$    & F1${_{ow}}$
\\ \cmidrule(r){1-7} 
{DF\cite{sirinam2018deep}}
& 38.44 & 36.25
& 51.15 & 55.76
& 63.47 & 71.14 
\\ \cmidrule(r){1-1} \cmidrule(r){2-3} \cmidrule(r){4-5} \cmidrule(r){6-7}
{FlowPrint\cite{van2020flowprint}}
& \textcolor{white}{x}79.65\textcolor{white}{x} & \textcolor{white}{x}80.04\textcolor{white}{x} 
& \textcolor{white}{x}68.79\textcolor{white}{x} & \textcolor{white}{x}62.36\textcolor{white}{x} 
& \textcolor{white}{x}58.29\textcolor{white}{x} & \textcolor{white}{x}69.01\textcolor{white}{x}
\\ \cmidrule(r){1-1} \cmidrule(r){2-3} \cmidrule(r){4-5} \cmidrule(r){6-7}
{PT-OM\cite{bendale2016towards}}
& 90.15 & 92.26
& 75.51 & 73.75 
& 83.07 & 89.01
\\ \cmidrule(r){1-1} \cmidrule(r){2-3} \cmidrule(r){4-5} \cmidrule(r){6-7}
{PT-ST\cite{yoshihashi2019classification}}
& 87.77 & 92.44 
& 72.33 & 73.19 
& 76.68 & 87.18 
\\ \cmidrule(r){1-1} \cmidrule(r){2-3} \cmidrule(r){4-5} \cmidrule(r){6-7}
\textbf{OWCP}
& \textbf{93.01} & \textbf{94.62}
& \textbf{80.51} & \textbf{77.85} 
& \textbf{88.69} & \textbf{90.96} 
\\ \bottomrule
\end{tabu}}
\end{table}

We perform closed-world and open-world scenario experiments on three datasets according to the comparison method adaptation, respectively.
As can be seen from Table \ref{Tab2} (focusing on closed-world) and Table \ref{Tab3} (focusing on open-world), OWCP outperforms all methods. In the closed-world scenario, our model achieves 3.85\% and 7.06\% improvement in F1 from existing methods, \emph{e.g.,} FlowPrint and PERT, on CrossPlatform, respectively, indicating that OWCP achieves strong feature representation through contrastive pre-training learning. 
Meanwhile, in the open-world scenario, OWCP achieves up to 14.58\% improvement in F1${{}_{ow}}$ over the best open-world ETC model, FlowPrint, on CrossPlatform. Our recognition paradigm also improves by 2.18\% and 2.36\% in F1${{}_{ow}}$ compared to two generic open-world solutions. 
Furthermore, by performing a more comprehensive analysis of the two tables, OWCP only declines by 1.17\% on ISCX17 and 1.59\% on CrossPlatform when challenged with unknown apps, proving that our model is adequate to combat the unknowns that arise in the real network environments.

\subsection{Ablation Study}
We present ablation results in Table \ref{Tab4} to evaluate the contribution of each component on the widely compared CrossPlatform. 
NP and PL refer to the non-plaintext payload and packet length sequences, respectively, and are used to assess the impact of different featur
es. The decrease of 2.17\%$\downarrow$ and 0.87\%$\downarrow$ on F1, and 2.75\%$\downarrow$ and 1.25\%$\downarrow$ on F1${_{os}}$, for the w/o NP and w/o PL models, respectively, suggests that both sequence features are beneficial in classification. Additionally, the effect of NP is superior to that of PL on the CrossPlatform.
CPT and BSF denote the contrastive pre-training and background similarity filtering, respectively. 
We remove the pre-trained model to evaluate the impact of contrastive pre-training. 
According to the w/o CPT model, the loss of pre-training damps the classification effect, especially 3.03\%$\downarrow$ drop in F1${_{os}}$ of open-world scenarios, indicating that discovering marginal flows without scene knowledge is unreliable and leads to unknown recognition even worse than thresholding Softmax (0.85\%$\downarrow$).
We also evaluated the effect of the model without background similarity filtering (w/o BSF) and found that not filtering homogeneous marginal flows had a more significant impact on known classes. In the closed-world scenario, F1 decreased by 1.15\%, indicating that using homogeneous marginal flows to simulate unknown ones will lead to misjudgment of  homogeneous flows of known classes.

\begin{table}[!htp]
\centering
\scriptsize
\captionsetup{font={footnotesize}}
\caption{Ablation Study of Key Components in OWCP on the CrossPlatform dataset. $\downarrow$ indicates the degree of decline from the full model.}
\label{Tab4}
\renewcommand\arraystretch{0.8}
\resizebox{\linewidth}{!}{
\begin{tabu}{l|cccc}
\toprule
\multirow{2}{*}{\textbf{Method}}   
& \multicolumn{2}{c}{{Closed-world}} 
& \multicolumn{2}{c}{{Open-world}} 
\\ \cmidrule(r){2-3} \cmidrule(r){4-5}
& AC 
& F1
& AC${}_{ow}$ 
& F1${}_{ow}$
\\ \cmidrule(r){1-5} 
w/o NP
& 96.76/\textcolor{black}{2.12$\downarrow$}
& 94.04/\textcolor{black}{2.17$\downarrow$}
& 89.35/\textcolor{black}{3.66$\downarrow$}
& 91.87/\textcolor{black}{2.75$\downarrow$}
\\ \cmidrule(r){1-1} \cmidrule(r){2-3} \cmidrule(r){4-5}
w/o PL
& 97.61/\textcolor{black}{1.27$\downarrow$}
& 95.34/\textcolor{black}{0.87$\downarrow$}
& 92.07/\textcolor{black}{0.94$\downarrow$}
& 93.37/\textcolor{black}{1.25$\downarrow$}
\\ \cmidrule(r){1-1} \cmidrule(r){2-3} \cmidrule(r){4-5}
w/o CPT
& 98.23/\textcolor{black}{0.65$\downarrow$}
& 92.34/\textcolor{black}{3.87$\downarrow$}
& 90.97/\textcolor{black}{2.04$\downarrow$}
& 91.59/\textcolor{black}{3.03$\downarrow$}
\\ \cmidrule(r){1-1} \cmidrule(r){2-3} \cmidrule(r){4-5}
w/o BSF
& 97.32/\textcolor{black}{1.56$\downarrow$}
& 95.06/\textcolor{black}{1.15$\downarrow$}
& 92.26/\textcolor{black}{0.75$\downarrow$}
& 93.81/\textcolor{black}{0.81$\downarrow$}
\\ \cmidrule(r){1-1} \cmidrule(r){2-3} \cmidrule(r){4-5}
\textbf{OWCP(full model)}
& \textcolor{white}{xll} 98.88\textcolor{white}{xll} 
& \textcolor{white}{xll} 96.21\textcolor{white}{xll} 
& \textcolor{white}{xll} 93.01\textcolor{white}{xll} 
& \textcolor{white}{xll} 94.62\textcolor{white}{xll} 
\\ \bottomrule
\end{tabu}}
\end{table}

\subsection{Sensitivity Analysis}

\begin{figure}[htbp]
\centering
\subfloat[]{\includegraphics[scale=0.3]{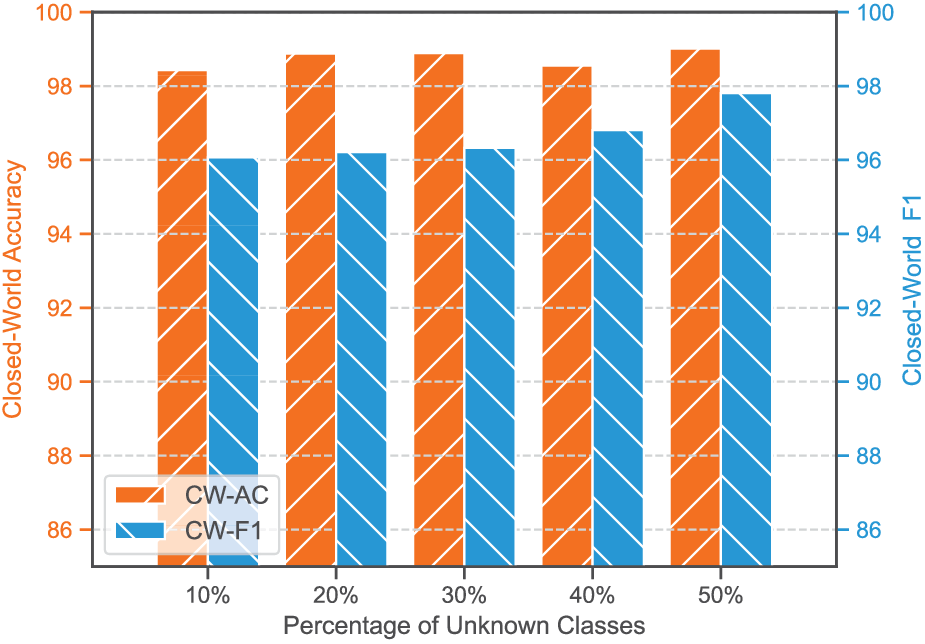}%
\label{fig7_first_case}}
\hspace{0.2cm}
\subfloat[]{\includegraphics[scale=0.3]{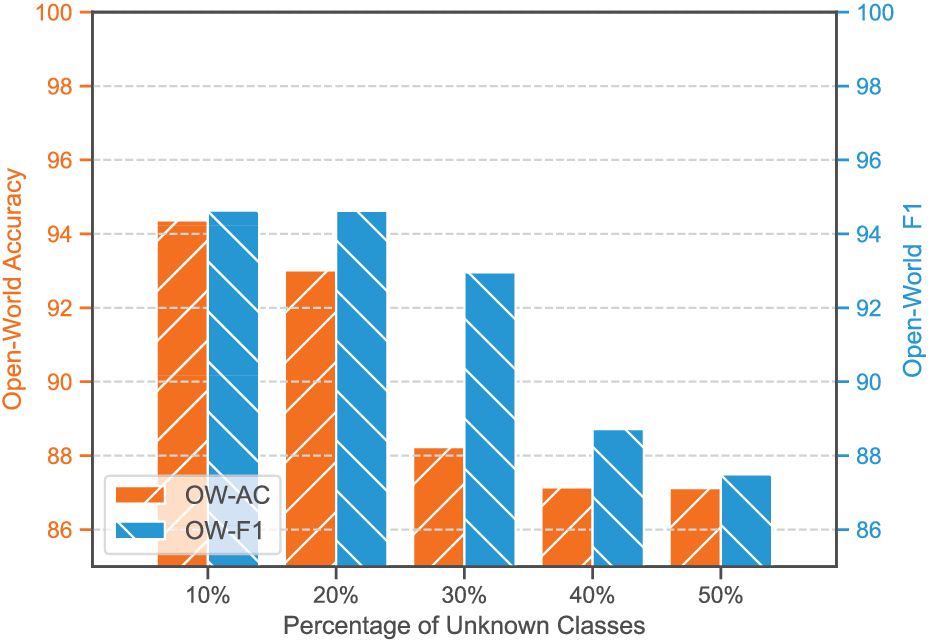}%
\label{fig7_second_case}}
\caption{Performance of Different Percentages of Unknown Classes on the CrossPlatform Dataset. (a) The Closed-World (CW) Metrics. (b) The Open-World (OW) Metrics.}
\label{fig5}
\end{figure}

We also conduct a sensitivity analysis to investigate the effect of the percentage of unknown classes on CrossPlatform. We sequentially select the percentage of unknown classes from [10\%, 20\%, 30\%, 40\%, 50\%] to observe the variation of closed-world and open-world metrics, as shown in Fig. \ref{fig5}. As the percentage of unknown classes increases, there is a decreasing trend in F1${}_{ow}$ in open-world scenarios and an increasing trend in F1 in closed-world scenarios. However, the gain from the known classes decline is more negligible than the loss in the open-world scenarios, which is the same as our real network environment perception. The percentage of unknown classes should be kept within a reasonable range, and once it exceeds 30\% will have a non-negligible impact.

\section{Conclusion}
In this paper, we propose a novel Open-World Contrastive Pre-training (OWCP) framework for ETC, which can effectively classify the known apps and detect the unknowns simultaneously. OWCP provides a robust feature representation by traffic contrastive pre-training. On this basis, we determine the spherical mapping space for each known class to find non-homogeneous flows at the margins, which are used to train GANs to synthesize new parts close to the known but not belonging to any class as the unknown flows. The synthetic flows are then assigned to unknown nodes of Softmax to modify the classifier, to overall reduce and flatten the recognition probability of unknown flows, making the classifier more sensitive to known flows and significantly suppressing unknown ones.
Our proposed method is evaluated in extensive experiments in three public available datasets, which shows that OWCP outperforms existing closed-world/open-world ETC methods and the generic open-world classification methods, with F1 and F1${}_{ow}$ achieving 96.21\% and 94.62\%. Furthermore, extensive ablation studies prove that the payload/packet length sequences, contrastive pre-training, and background similarity filtering significantly contribute to the performance improvements of OWCP. Meanwhile, the sensitivity analysis also showed that the percentage of unknown classes should be kept within a reasonable range. 
Once it exceeds 30\% will have a non-negligible impact.


\end{document}